\documentclass{mn2e}
\usepackage{graphicx}
\def\simgt{\ {\raise-.5ex\hbox{$\buildrel>\over\sim$}}\ }

\begin{document}

\title{Asteroseismological studies of three $\beta$ Cephei stars: IL Vel,
V433~Car and KZ Mus}
\author[Handler et al.]
    {G. Handler$^{1,2}$, R. R. Shobbrook$^{3}$, F. F. Vuthela$^{2,4}$,
L. A. Balona$^{2}$, F. Rodler$^{1}$, \and T. Tshenye$^{4}$
	\and \\
$^{1}$ Institut f\"ur Astronomie, Universit\"at Wien,
T\"urkenschanzstra\ss e 17, A-1180 Wien, Austria\\
$^{2}$ South African Astronomical Observatory, P.O. Box 9, Observatory
7935, South Africa\\
$^{3}$ Visiting Fellow, Australian National University, Canberra, ACT,
Australia\\
$^{4}$ Department of Physics, University of the North-West,
Private Bag X2046, Mmabatho 2735, South Africa
}

\date{Accepted 2003 nnnn nn.
   Received 2002 nnnn nn;
   in original form 2002 nnnn nn}

\maketitle

\begin{abstract} 

We have acquired between 127 and 150 h of time-resolved multicolour
photometry for each of the three $\beta$ Cephei stars IL Vel, V433 Car and
KZ Mus over a time span of four months from two observatories. All three
objects are multiperiodic with at least three modes of pulsation.

Mode identification from the relative colour amplitudes is performed. We
obtain unambiguous results for the two highest-amplitude modes of IL Vel
(both are $\ell=1$) and the three strongest modes of KZ Mus ($\ell=2,0$
and 1), but none for V433 Car. Spectroscopy shows the latter star to be a
fast rotator ($v \sin i=240$\,km/s), whereas the other two have moderate
$v \sin i$ (65 and 47 km/s, respectively).

We performed model calculations with the Warsaw-New Jersey stellar
evolution and pulsation code. We find that IL Vel is an object of about 12
$M_{\sun}$ in the second half of its main sequence evolutionary track. Its
two dipole modes are most likely rotationally split components of the mode
originating as $p_1$ on the ZAMS; one of these modes is $m=0$. V433 Car is
suggested to be an unevolved 13 $M_{\sun}$ star just entering the $\beta$
Cephei instability strip. KZ Mus seems less massive ($\approx 12.7
M_{\sun}$) and somewhat more evolved, and its radial mode is probably the
fundamental one. In this case its quadrupole mode would be the one
originating as $g_1$, and its dipole mode would be $p_1$.

Two of our photometric comparison stars also turned out to be variable. HD
90434 is probably a new slowly pulsating B star whose dominant mode is a
dipole, whereas the variability of HD 89768 seems to be due to binarity.

It is suggested that mode identification of slowly rotating $\beta$ Cephei
stars based on photometric colour amplitudes is reliable; we estimate that
a relative accuracy of 3\% in the amplitudes is sufficient for unambiguous
identifications. Due to the good agreement of our theoretical and
observational results we conclude that the prospects for asteroseismology
of multiperiodic slowly rotating $\beta$ Cephei star are good.

\end{abstract}


\section{Introduction}

The $\beta$ Cephei stars are a group of early B-type stars of luminosity
classes III-V that pulsate in pressure (p) and gravity (g) modes of low
radial overtone. Their pulsational behaviour is thus similar to that of
the A/F-type $\delta$ Scuti stars. Both classes of variable seem suitable
for asteroseismological investigations: their pulsational frequencies may
be used to sound their interiors as the associated modes penetrate deeply
into the star.

As exciting as this possibility is in theory, practice has shown that
several obstacles still need to be overcome before precision
asteroseismology of $\beta$ Cephei or $\delta$ Scuti stars can be done. On
the theoretical side an adequate description of higher-order rotational
effects is being worked on (e.g. Soufi, Goupil \& Dziembowski 1998), and
mode coupling is expected to have major effects on the pulsational
eigenspectra of fast rotators (see Daszy{\'n}ska-Daszkiewicz et al. 2002).
Observationally, the main problem is the detection and identification of
as many pulsation modes as possible because a complete set of mode spectra
has not even nearly been detected for any of these pulsators.

It may be suspected that the $\delta$ Scuti stars offer better
possibilities for asteroseismology than $\beta$ Cephei stars do, as they
generally have many more pulsation modes excited to observable amplitudes.
This result is in agreement with theoretical investigations of mode
excitation (e.g.\,Pamyatnykh 2003). However, the problem lies with mode
identification, that is, the match of the observed frequencies with the
corresponding pulsation mode defined by the three parameters $k$, the
radial overtone of the mode, the spherical degree $\ell$ and the azimuthal
order $m$.

To accomplish such a match in the presence of incomplete observed
pulsation spectra, it is necessary to use mode identification methods.
However, the surface convection zones of $\delta$ Scuti stars are expected
to affect some of these methods, like the commonly used photometric method
that utilises and amplitude ratios and/or phase shifts of the light curves
in different filters, adversely (see Balona \& Evers 1999), and it is not
known if they can be trusted. The $\beta$ Cephei stars do not have surface
convection zones and therefore the simple photometric method may be
applicable with more confidence.

We have therefore chosen three $\beta$ Cephei stars whose literature data
qualify them as interesting candidates for in-depth studies aiming at the
detection of many pulsation frequencies and at their mode identification
by photometric means. All these objects have been reported as
multiperiodic variables before and photometric mode identifications have
been attempted. However, in all cases the amount of data available was
small enough to give us hope for a significant improvement over previous
results if we can obtain $\simgt 100$ h of observation.

The most comprehensive photometric study of IL Vel (HD 80383) was
performed by Heynderickx \& Haug (1994). They found four frequencies in
their UBV light curves, three of which formed an equally spaced triplet,
suggestive of rotational $m$-mode splitting. This frequency solution did
however not fit additional data in the Walraven system within the
precision of the measurements. Heynderickx, Waelkens \& Smeyers
(1994) used the Walraven data for mode identification and suggested that
all modes are of degree $\ell=0$ and $\ell=1$.

The variability of V433 Car (HD 90288) was discovered by Lampens (1988)
and followed up by Heynderickx (1992). The frequency analysis of all the
Geneva and Walraven data available to Heynderickx (1992) allowed him to
disentangle four frequencies in the light variations of V433 Car. The mode
identification by Heynderickx et al. (1994) was somewhat ambiguous, but
suggested a mixture of (possibly) radial and nonradial modes with
spherical degrees $\ell$ up to 4.

Our third target star, KZ Mus (HD 109885) was a discovery of the HIPPARCOS
satellite (ESA 1997) reported by Waelkens et al. (1998). Aerts
(2000) obtained new Geneva photometry for the star and performed both a
frequency analysis of all data and mode identification. She suggested that
the star pulsates with at least three frequencies, where the second
strongest one appeared to be radial.

\section{Observations}

\subsection{Photometry}

We have studied IL Vel, V433 Car and KZ Mus from both the Siding Spring
(SSO, Australia) and South African Astronomical (SAAO) Observatories. We
used photoelectric photometers attached to the 0.6-m telescope at SSO and
the 0.5-m and 0.75-m telescopes at SAAO. 

The most valuable part of the energy distribution of a $\beta$ Cephei star
for mode identification by means of photometry is in the blue ($\lambda <
4200$~\AA). Therefore, the Geneva and (particularly) Walraven systems are
well suited for such a purpose as they have several filters in this
wavelength domain. Unfortunately, these are not widely available. Owing to
the faintness of our variables ($V=8.1-9.1$) we decided to utilise the
standard Johnson UBV system, supplemented by Str{\o}mgren v whenever
possible.

The periods of the known pulsations of our three targets are all longer
than 150 minutes, and the stars are all located in an area covering some
30\degr on the sky. Consequently, we chose one local comparison star for
each variable (the B5 IV star HD 79670 for IL Vel, the B9 IV/V star HD
90434 that was later replaced by the A1 IV/V star HD 89768 for V433 Car
and the B9 IV star HD 109082 for KZ Mus; all spectral types are from Houk
\& Cowley 1975), and our observing sequence included all six stars
whenever reachable. The resulting cycle time of about 25 minutes (or less)
per measurement of each variable sampled the light curves of all stars
adequately.

Another aspect of the pulsations of $\beta$ Cephei stars requires
consideration when planning an observational effort such as ours. In many
cases, closely spaced pulsation frequencies are present. As mentioned
above Heynderickx \& Haug (1994) reported just that for our target IL Vel.
Consequently, a sufficient time baseline must be spanned by the
observations (at least 2 months for IL Vel) for such frequencies to be
resolved and the measurements should be distributed evenly over the
observation period to avoid aliasing problems.

\begin{figure*}
\includegraphics[width=185mm,viewport=5 00 512 490]{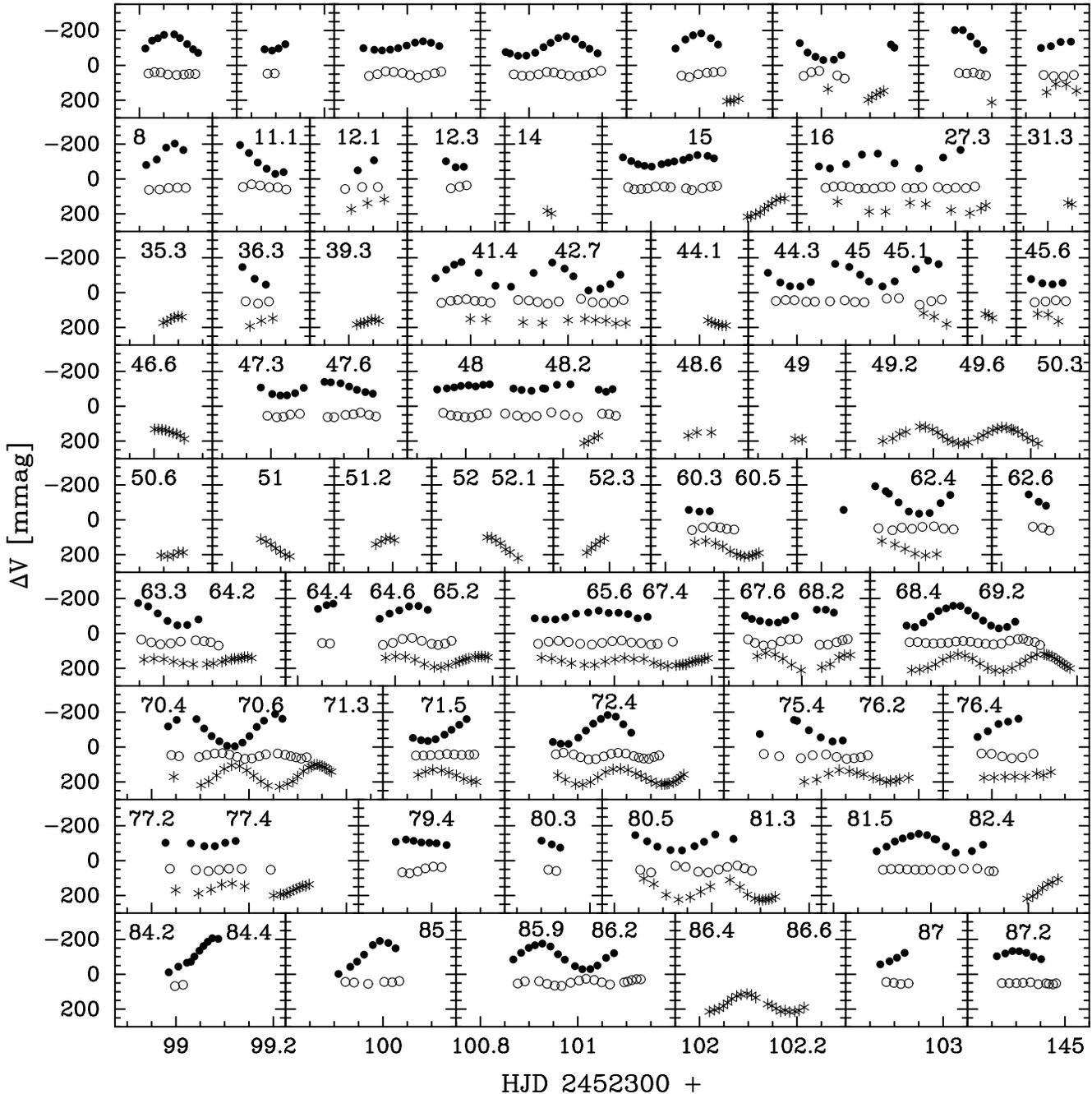}
\caption[]{V-filter light curves from all our measurements of IL Vel
(filled circles), V433 Car (open circles) and KZ Mus (asterisks). Note the
shorter periods and much lower amplitude of V433 Car. Multiperiodicity of
all three stars is evident.}
\end{figure*}

We have therefore also observed these objects during a multisite campaign
for the $\delta$ Scuti star FG Vir (Breger et al., in preparation) when
this star was not reachable. This only resulted in short runs up to two
hours that however covered important gaps in the total coverage which made
them quite valuable. The main body of our final data set spanned 94 days
in February to May 2002, during which data was obtained on 52 nights. A
total of 146 hr of measurement of IL Vel, 150 hr of V433 Car and 127 hr of
KZ Mus was acquired.

Data reduction was performed in the standard way for differential
photoelectric time-series photometry. First, we corrected the data for
coincidence losses, followed by sky subtraction and by extinction
determination using the time series of the constant stars in our ensemble.
After the extinction correction, the differential light curves of the
variables with respect to their local comparison stars and that of the
three comparison with respect to the others were calculated and the times
of measurement were converted to Heliocentric Julian Date (HJD). The light
curves of the three $\beta$ Cephei stars in the V filter are shown in
Fig.\,1.

As part of the reductions were already performed during the observational
campaign, it was noticed that the comparison star HD 90434 was variable.
Consequently, it was replaced by HD 89768. As it turned out during the
final reductions after completion of the observations, this star is also
variable. We will analyse these objects later, but for the time being let
it suffice to say that we performed a frequency analysis of their
differential light curves with respect to the constant comparison stars
for the other targets. We then subtracted synthetic light curves built
from the results of this frequency search from the differential
measurements of V433 Car so the variability of its comparison stars will
not affect the frequency analysis of this star. This procedure is valid as
the time scales of the variability of HD 90434 and HD 89768 is much longer
than that of V433 Car.

\subsection{Spectroscopy}

With the main aim of determining the projected rotational velocities of IL
Vel, V433 Car and KZ Mus, high-resolution spectra of the three $\beta$
Cephei stars were obtained on the night of 4/5 May 2002 with the 1.9-m
telescope at SAAO. We used the {\tt GIRAFFE} echelle fibre-fed
spectrograph attached to the Cassegrain focus of this telescope. The {\tt
GIRAFFE} spectrograph has a resolving power of about 32\,000, giving a
resolution of 0.06 -- 0.09~{\AA} per pixel. Exposure times were 1500
seconds for IL Vel and KZ Mus as well as 600 seconds for V433 Car for 7S/N
ratios between 27 and 35. A Th-Ar arc lamp was used for wavelength
calibration.

The spectra were normalised to the continuum by using an unbroadened
synthetic spectrum with T$_{\rm eff} = 23000$~K, $\log g = 4.00$ as a
template, using the {\tt SPECTRUM} code (Gray \& Corbally 1994). A running
median of each echelle order was divided by the corresponding section of
the synthetic spectrum and taken to represent the response function of the
instrument. A polynomial of degree 5 was fitted to the response function
and used to correct the observed spectrum.

For each order, the observed rectified spectrum was correlated with the
corresponding section of the synthetic spectrum after removing the unit
continuum, effectively resulting in the mean line profile with the
continuum removed. A quadratic was then fitted to the correlation
function, and its maximum adopted as the ``radial velocity'' for each
order. The mean radial velocity from all the orders is also obtained.

The correlation profiles can be co-added to form what is essentially a
mean line profile. The projected rotational velocity, $v \sin i$, can be
determined by fitting a model profile of a rotating star.  The projected
rotational velocity is adjusted until a best fit is obtained to the
observed profile. We show these fits in Fig.\,2; the results from our
spectroscopic analysis are summarised in Table 1.

\begin{figure}
\includegraphics[width=99mm,viewport=00 00 625 580]{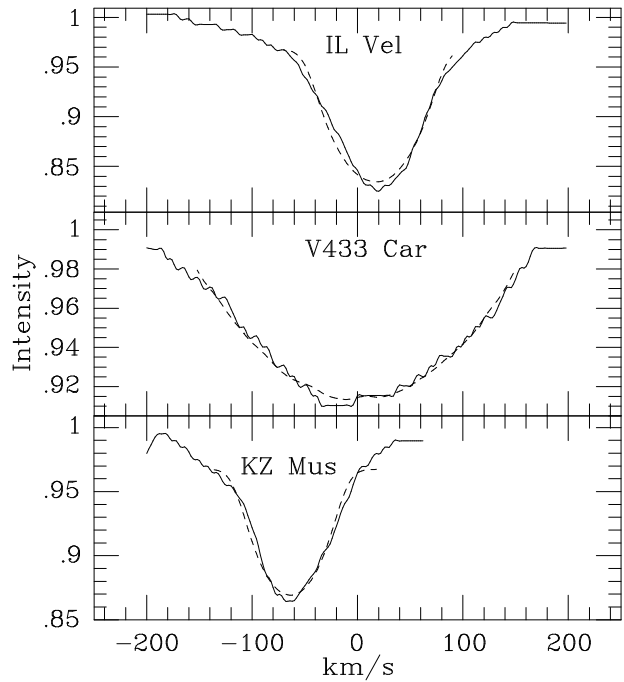}
\caption[]{The average Helium line profile for our three variables (full
line) overlaid by a model atmosphere fit (dashed lines) to determine the
projected rotational velocities. The line profile asymmetries are likely
caused by nonradial stellar pulsation.}
\end{figure}

\begin{table}
\caption[]{Results of our determinations of the radial and projected
rotational velocities.}
\begin{flushleft}
\begin{tabular}{lcc}
\hline
Star & $v \sin i$ [km/s] & $<Vr>$ [km/s] \\
\hline
IL Vel & 65 $\pm$ 3 & +19.0 \\
V433 Car & 240 $\pm$ 10 & +4.0 \\
KZ Mus & 47 $\pm$ 3 & -61.6 \\
\hline
\end{tabular}
\end{flushleft}
\end{table}

\section{Frequency analysis of the photometric time series}

Our frequency analysis was performed with the program {\tt PERIOD 98}
(Sperl 1998). This package applies single-frequency power spectrum
analysis and simultaneous multi-frequency sine-wave fitting. One of its
many advanced features is its capability to fix dependent signal
frequencies to certain values, e.g. to sum and difference terms, and to
perform simultaneous nonlinear least squares fits with such fixed
frequencies.

Our strategy for the frequency analysis starts with the calculation of the
spectral window of the data. It is computed as the Fourier amplitude
spectrum of a single noise-free sine wave with the frequency of highest
amplitude in our data. In this way, the reflection of the spectral window
pattern at zero frequency, which may not be negligible as our observing
nights often did not sample a full cycle of the light variations, can be
evaluated.

We continue by computing amplitude spectra of our data as well as those of
residual light curves after the previously identified periodicities had
been removed using the multi-periodic fitting algorithm of {\tt PERIOD
98}. We continue this process until no significant peaks are left in the
residual amplitude spectrum. We consider an independent peak significant
if it exceeds four times the local noise level in the amplitude spectrum,
following Breger et al.\,(1993). Combination signals only require
$S/N>3.5$ to be regarded as significant. Experience has shown that this
criterion is both reliable and conservative.

Similar analyses were performed for all the four filters used. We note
that, within the uncertainties of frequency determination, these detected
frequencies were the same for all stars in every filter. This confirms
that alias ambiguities do not affect our analysis. As our final values for
the frequencies we adopted the mean values from the data in all four
filters weighted by the associated signal-to-noise ratios. The deviations
of the optimum frequencies in the individual filters from this mean is
taken as our error estimate on the frequency determination.

\subsection{KZ Mus}

As the frequency analysis of this star was straightforward, it is the
first that we chose to describe. The spectral window, amplitude spectra,
and prewhitened versions thereof are shown in Fig.\,3. We use the data
obtained in the B filter for display as they have best S/N ratio. Due to
our two-site coverage the spectral window is sufficiently clean so that
alias ambiguities do not represent a problem for the analysis.

\begin{figure}
\includegraphics[width=99mm,viewport=5 02 305 562]{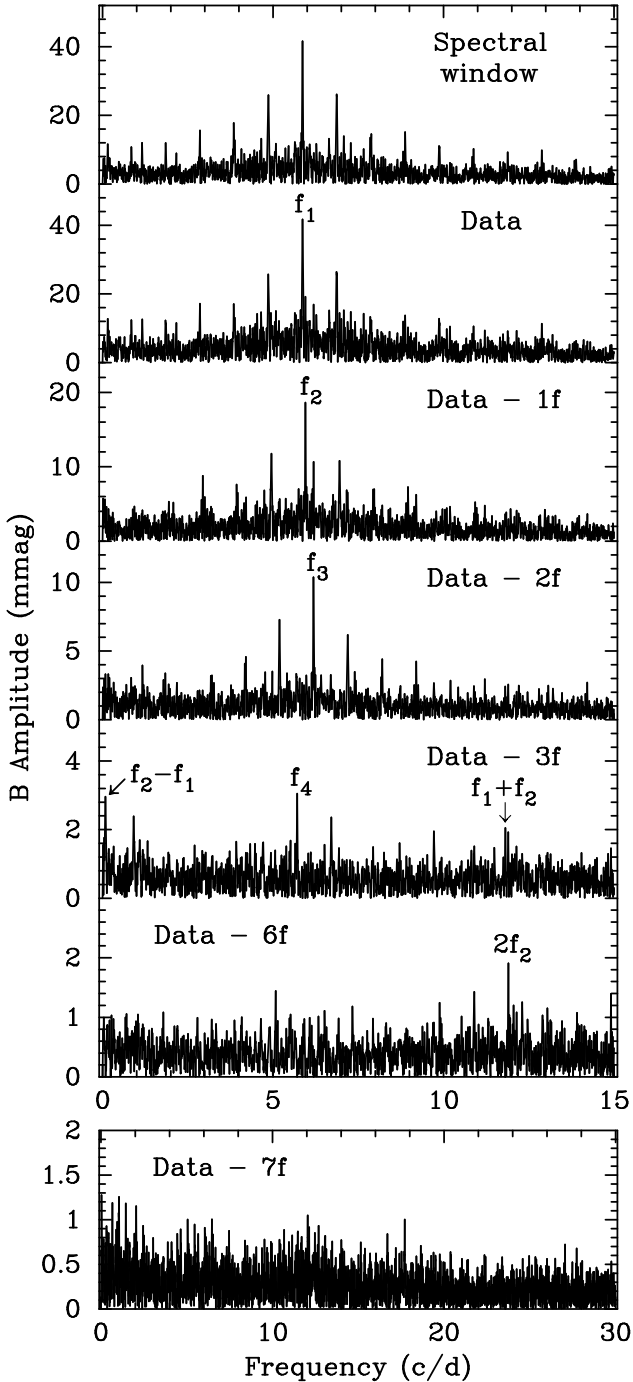}
\caption[]{Spectral window and amplitude spectra of our B filter data of
KZ Mus. Some prewhitening steps of the detected frequencies are shown in
consecutive panels. The lowest panel shows the amplitude spectrum of the
combined UvBV residuals after all significant signals have been removed
from the measurements.}
\end{figure}

The first three signals in the periodogram are easy to detect. After
prewhitening these, a more complicated residual amplitude spectrum (fifth
panel of Fig.\,3), with several peaks standing out, remains. A closer look
reveals that two of them are the combination sum and difference
frequencies of the two strongest modes, whereas the remaining tall peak
corresponds to an independent signal. Prewhitening all the six variations
detected so far, the 2f-harmonic of the second strongest mode also becomes
notable.

After this signal is included in our multifrequency light-curve fit, no
significant peaks remain in the residual amplitude spectrum of the B
filter data alone, and neither are there any present in the measurements
in any other filter. However, we can attempt to increase the signal to
noise in our analysis by adding the data in all filters. Obviously, this
must not be done for the original light curves, as different pulsation
modes have different relative amplitudes in the different wavelength
passbands (that are a diagnostic for mode identification!). It is however
safe to combine the residual light curves as we are working at low S/N
ratio, i.e. the intrinsic differences in mode amplitude between the
wavebands are small compared to the noise in the data.

To combine the different residual data sets we have therefore subtracted a
multifrequency solution from all the individual filters' observations,
using exactly the same frequencies (determined as described in the
previous section) for each waveband. We then roughly scaled the residuals
to B amplitude by multiplying the U filter residuals by 0.8 and the V
filter residuals by 1.1. We did not scale the v filter data as the signals
in it were found to have essentially the same amplitude as in B. The scale
factors are approximate average amplitude ratios of the modes previously
detected.

All these residual light curves were then added together, averaged, and
the periodogram was computed out to the Nyquist frequency (lowest panel of
Fig.\,3); we find no more significant signals. Interestingly, we do note
two residual mounds of amplitude, one in the frequency region where the
pulsational signals were found and another one where combination
frequencies would be expected. In fact, the two highest peaks in the
latter region are located at exact sums of two known frequencies but they
are not formally significant.

We can therefore claim the detection of seven signals in the light curves
of KZ Mus; our final frequency solution is listed in Table 2. The rms
scatter per single data point in the combined and averaged UvBV residuals
is 3.0 mmag.

\begin{table*}
\caption[]{Multifrequency solution for our KZ Mus data. Frequencies are
mean values from results of all four filters weighted by the S/N ratio of 
the signal. Errors on the amplitudes were calculated with the formulae of
Montgomery \& O'Donoghue (1999). The S/N ratio quoted is the average of 
the values in the individual filters.}
\begin{flushleft}
\begin{tabular}{ccccccc}
\hline
ID & Frequency & \multicolumn{4}{c}{Amplitude} & mean S/N\\
 & (c/d) & U (mmag) & B (mmag) & V (mmag) & v (mmag) & \\
\hline
$f_1$ & 5.86384 $\pm$ 0.00006 & 45.6 $\pm$ 0.3 & 40.9 $\pm$ 0.3 & 38.5 $\pm$ 0.3 & 41.2 $\pm$ 0.4 & 89.4 \\
$f_2$ & 5.95026 $\pm$ 0.00011 & 28.8 $\pm$ 0.3 & 19.6 $\pm$ 0.3 & 16.4 $\pm$ 0.3 & 20.6 $\pm$ 0.4 & 45.6 \\
$f_3$ & 6.1874 $\pm$ 0.0005 & 14.3 $\pm$ 0.3 & 11.0 $\pm$ 0.3 & 10.5 $\pm$ 0.3 & 11.8 $\pm$ 0.4 & 25.7\\
$f_4$ & 5.7090 $\pm$ 0.0024 & 3.4 $\pm$ 0.3 & 3.2 $\pm$ 0.3 & 2.4 $\pm$ 0.3 & 3.1 $\pm$ 0.4 & 6.6\\
$f_2-f_1$ & 0.08642 & 4.3 $\pm$ 0.3 & 2.8 $\pm$ 0.3 & 3.1 $\pm$ 0.3 & 3.1 $\pm$ 0.4 & 6.3\\
$f_1+f_2$ & 11.81410 & 2.5 $\pm$ 0.3 & 2.2 $\pm$ 0.3 & 2.5 $\pm$ 0.3 & 2.6 $\pm$ 0.4 & 5.7\\
$2f_2$ & 11.90052 & 2.6 $\pm$ 0.3 & 1.9 $\pm$ 0.3 & 2.0 $\pm$ 0.3 & 2.6 $\pm$ 0.4 & 5.2\\
\hline
\end{tabular}
\end{flushleft}
\end{table*}

\subsubsection{Re-analysis of published data}

Our frequency solution in Table 2 differs somewhat from the results of
Aerts (2000), who needed to base her frequency analysis on the HIPPARCOS
data of the star. The first two signals detected by her agree with ours,
but the third one is different; the remaining signals are below her
detection level. Two possible reasons for the disagreement can be
imagined, the more exciting one being intrinsic amplitude variability of
the star and the other one being an alias problem in the poorly sampled
HIPPARCOS photometry.

Consequently, we have re-analysed the time series of KZ Mus acquired by
the HIPPARCOS satellite. The residual amplitude spectrum after
prewhitening the two consistent main frequencies contains the third signal
claimed by Aerts (2000) as the tallest peak. However, our third signal is
also visible and of similar strength.  As our data are more extensive and
of better quality than the HIPPARCOS photometry, we suggest that the third
signal found by Aerts (2000) is an unfortunate artifact of the poor
sampling of the HIPPARCOS measurements.

\subsection{V433 Car}

The upper panel of Fig.\,4 shows that the spectral window function for our
measurements of this star is also quite good. The amplitude spectrum is
dominated by two peaks of similar strength that seem to imply an aliasing
problem, but thanks to the long time base of our data they are well
resolved and do not interfere with each other; they are simply two
independent signals of similar amplitude.

\begin{figure}
\includegraphics[width=99mm,viewport=5 02 305 490]{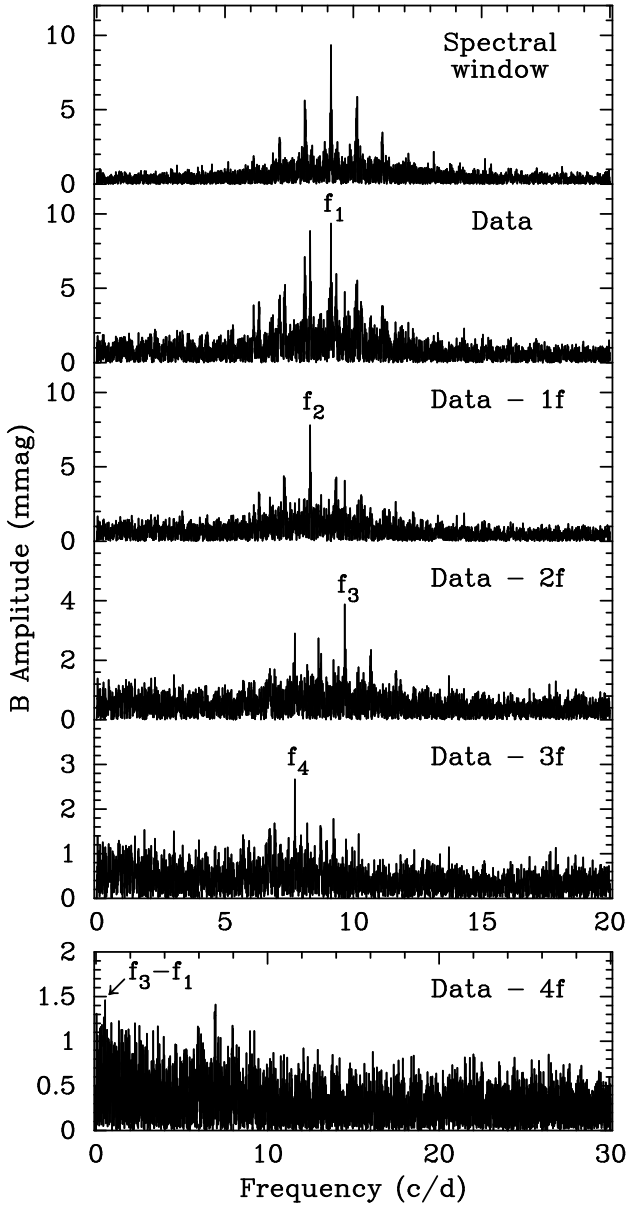}
\caption[]{Spectral window and amplitude spectra of our B filter data of
V433 Car. Some prewhitening steps of the detected frequencies are shown in
consecutive panels. The lowest panel shows the amplitude spectrum of the
combined UvBV residuals after all previously significant signals have been 
removed from the measurements. One combination frequency is detected that way.}
\end{figure}

Prewhitening of these variations results in the detection of a third
frequency and the next step reveals another significant peak. After these
four periodicities have been removed, no obvious new candidate frequency
can be discerned. We therefore combined the residual data as we did for KZ
Mus, but used somewhat different scale factors that seemed more
appropriate for this star. We multiplied the U residuals by 0.83 and the
ones in V by 1.05 to obtain approximate B amplitudes.

The amplitude spectrum of the residuals thus combined is shown in the
lowest panel of Fig.\,4, again out to the Nyquist frequency. The highest
peak corresponds to a combination frequency difference and as it exceeds
the noise level more than 3.5 times, we adopt it as an additional term for
our frequency solution. We would also like to point out the peak near 7
c/d in the lowest panel of Fig.\,4. It is in fact a close doublet at 6.924
and 6.968 c/d that is present in the residuals of all individual filters,
but as it still does not exceed $S/N>4$, we cannot be sure about its
reality. We therefore adopt a five-frequency solution for our light curves
that we list in Table 3.

\begin{table*}
\caption[]{Multifrequency solution for our V433 Car data. Frequencies are
mean values from results of all four filters weighted by the S/N ratio of 
the signal. Errors on the amplitudes were calculated with the formulae of
Montgomery \& O'Donoghue (1999). The S/N ratio quoted is the average of
the values in the individual filters.}
\begin{flushleft}
\begin{tabular}{ccccccc}
\hline
ID & Frequency & \multicolumn{4}{c}{Amplitude} & mean S/N\\
 & (c/d) & U (mmag) & B (mmag) & V (mmag) & v (mmag) & \\
\hline
$f_1$ & 9.12909 $\pm$ 0.00014 & 9.9 $\pm$ 0.5 & 8.2 $\pm$ 0.3 & 8.2 $\pm$ 0.3 & 8.8 $\pm$ 0.5 & 17.9 \\
$f_2$ & 8.31649 $\pm$ 0.00009 & 10.0 $\pm$ 0.5 & 8.2 $\pm$ 0.3 & 7.3 $\pm$ 0.3 & 8.0 $\pm$ 0.5 & 16.7 \\
$f_3$ & 9.66707 $\pm$ 0.00024 & 4.3 $\pm$ 0.5 & 3.8 $\pm$ 0.3 & 4.0 $\pm$ 0.3 & 4.2 $\pm$ 0.5 & 8.5\\
$f_4$ & 7.7214 $\pm$ 0.0012 & 3.4 $\pm$ 0.5 & 2.8 $\pm$ 0.3 & 2.3 $\pm$ 0.3 & 2.6 $\pm$ 0.5 & 5.4\\
$f_3-f_1$ & 0.53798 & 1.8 $\pm$ 0.5 & 1.2 $\pm$ 0.3 & 1.5 $\pm$ 0.3 & 1.3 $\pm$ 0.5 & 3.7\\
\hline
\end{tabular}
\end{flushleft}
\end{table*}

The rms scatter of the combined residual light curves of V433 Car is 3.9
mmag per point. This is 30\% higher than for KZ Mus. Likely reasons for
the increased noise are undetected further intrinsic signals and possible
residual variations by the variable comparison stars that could not be
taken out.

\subsubsection{Comparison with literature data}

Heynderickx (1992) reported the detection of four frequencies in his light
curves of V433 Car. Two of them agree with the two strongest signals in
our Table 3, and the other two are 1 c/d aliases of our values. We prefer
the frequencies derived our analysis because our data are more numerous,
the results in the different filters were consistent and as our two-site
measurements have a much better spectral window function.

The order in which we detected the variation frequencies of V433 Car is
different from the one obtained by Heynderickx (1992). Although this
author does not list the amplitudes of the signals he determined,
inspection of his published PDM periodograms suggests that the pulsations
suffered amplitude variability in the 14 years that have elapsed between
his and our observations.

\subsection{IL Vel}

The spectral window and amplitude spectra of our measurements of IL Vel
are shown in Fig.\,5. Two independent frequencies dominate the
periodogram. After prewhitening those, a third signal stands out, and
including that one in the multifrequency solution allows the detection of
two combinations frequencies. Combining the residuals in all filters,
scaling the U data by a factor of 0.76 and the V data by 1.05, results in
the uninformative periodogram in the lowest panel of Fig.\,5. The
frequency solution for IL Vel is given in Table 4.

\begin{figure}
\includegraphics[width=99mm,viewport=5 00 305 490]{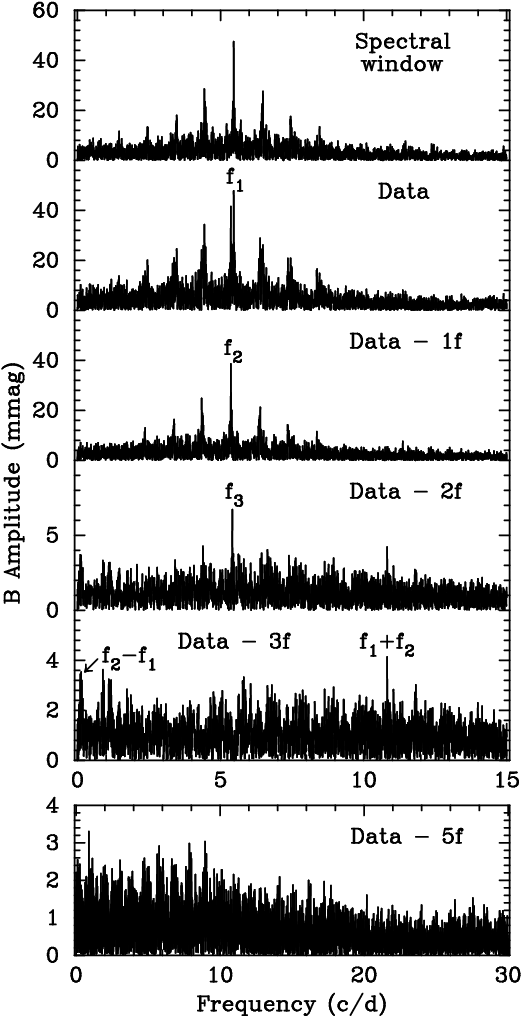}
\caption[]{Spectral window and amplitude spectra of our B filter data of
IL Vel. Some prewhitening steps of the detected frequencies are shown in
consecutive panels. The lowest panel shows the amplitude spectrum of the
combined UvBV residuals after all significant signals have been removed
from the measurements.}
\end{figure}

\begin{table*}
\caption[]{Multifrequency solution for our IL Vel data. Frequencies are
mean values from results of all four filters weighted by the S/N ratio of 
the signal. Errors on the amplitudes were calculated with the formulae of
Montgomery \& O'Donoghue (1999).The S/N ratio quoted is the average of the
values in the individual filters.}
\begin{flushleft}
\begin{tabular}{ccccccc}
\hline
ID & Frequency & \multicolumn{4}{c}{Amplitude} & mean S/N\\
 & (c/d) & U (mmag) & B (mmag) & V (mmag) & v (mmag) & \\
\hline
$f_1$ & 5.45976 $\pm$ 0.00012 & 59.9 $\pm$ 0.8 & 45.5 $\pm$ 0.6 & 43.2 $\pm$ 0.7 & 45.6 $\pm$ 0.7 & 43.2 \\
$f_2$ & 5.36293 $\pm$ 0.00011 & 53.0 $\pm$ 0.8 & 40.7 $\pm$ 0.6 & 38.3 $\pm$ 0.7 & 40.6 $\pm$ 0.7 & 38.4 \\
$f_3$ & 5.41340 $\pm$ 0.0010 & 8.7 $\pm$ 0.8 & 7.1 $\pm$ 0.6 & 6.5 $\pm$ 0.7 & 7.3 $\pm$ 0.7 & 6.6\\
$f_1+f_2$ & 10.82269 & 5.2 $\pm$ 0.8 & 4.0 $\pm$ 0.6 & 4.1 $\pm$ 0.7 & 3.7 $\pm$ 0.7 & 4.2\\
$f_2-f_1$ & 0.09683 & 5.3 $\pm$ 0.8 & 3.3 $\pm$ 0.6 & 3.4 $\pm$ 0.7 & 4.4 $\pm$ 0.7 & 3.7\\
\hline
\end{tabular}
\end{flushleft}
\end{table*}

This analysis may appear simple, but it is not. Firstly, the third
independent signal is located halfway between the first two that are of
much higher amplitude. Although it is well resolved from the two strong
modes, some doubts remain as to its reality; it could be an artifact
caused by amplitude/frequency variability of the two main
signals. 

We have therefore divided our data set into two halves (before and after
JD 2452360) and analysed them separately. We adopted the frequencies of
the two strongest modes and their detected combination signals as definite
and fitted them to each half, but left their amplitudes and phases as free
parameters. We then computed amplitude spectra of the residuals for each
half in each filter. The signal $f_3$ was detected in each of the 8
subsets of data; we conclude that it is real.

The second cause of concern is the rms residual of our five-frequency fit
to the IL Vel light curves. After combination of the data in all four
filters, it is an enormous 7.8 mmag per point, twice as large as for V433
Car and 2.6 times larger than for KZ Mus. The noise level in the lowest
panel of Fig.\,5 is also correspondingly higher.

We started our quest for the reason for the high residuals by
investigating the constancy of the comparison star. We calculated
differential light curves to all the other stars that we had simultaneous
data for, but found no evidence for variability of HD 79670, at least at a
level that could produce the high residuals observed. They must therefore
be (mostly) due to IL Vel.

When we checked the reality of $f_3$, we noted that its location in the
amplitude spectra of the two subsets of data was somewhat shifted and that
there was more structure surrounding it than just the normal spectral
window pattern plus noise. A plot of the residual light curves after
subtraction of our five-frequency solution shows conspicuous variability
on time scales similar to those of the known pulsational signals. We are
therefore inclined to think that further, presently unresolved, pulsation
frequencies are present in the light curves of IL Vel and that a data set
with an even longer time base than ours is required to detect them.

\subsubsection{Re-analysis of published data}

The results of our frequency analysis differ from that by Heynderickx \&
Haug (1994). Our first frequency is significantly different from theirs,
but we agree on the second one. Heynderickx \& Haug (1994) did not find
our third frequency, whereas we did not detect their third and fourth
signals.

To clarify the situation, we have retrieved the data by Heynderickx \&
Haug (1994) and re-analysed them. We show the periodogram analysis of
their Johnson B data in Fig.\,6, where we have assumed our frequencies to
be correct, and used them as the initial values for light curve-fitting
and prewhitening.

\begin{figure}
\includegraphics[width=99mm,viewport=5 00 305 299]{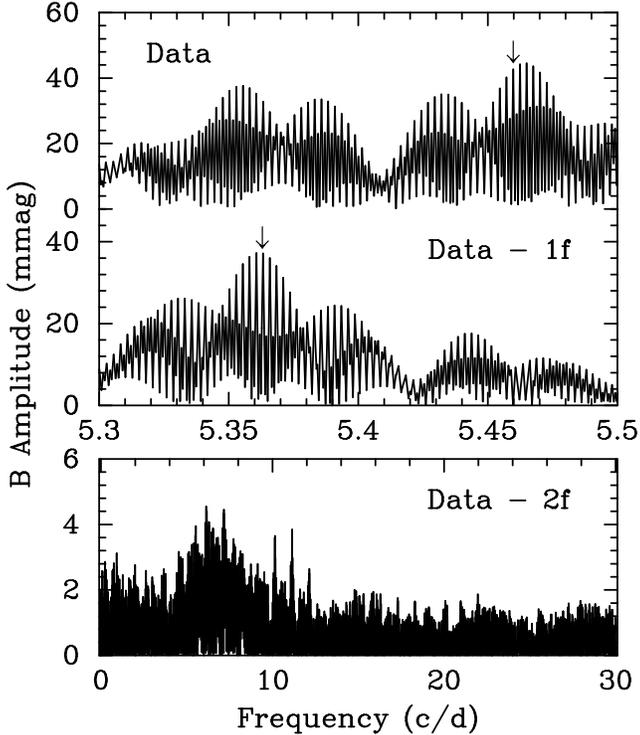}
\caption[]{Amplitude spectra of Heynderickx \& Haug's (1994) B filter data
of IL Vel; consecutive panels show some prewhitening steps. The positions
of our frequencies $f_1$ and $f_2$ are indicated with arrows and adopted
as definite for the prewhitening.}
\end{figure}

Our two frequencies $f_1$ and $f_2$ are sufficient to explain the dominant
variations in these measurements. After they are prewhitened, the residual
amplitude spectrum (lowest panel of Fig.\,6) shows no peak exceeding 5
mmag; the two additional frequencies claimed by Heynderickx \& Haug (1994)
had B amplitudes of 20 and 14 mmag, respectively. We conclude that these
were artifacts originating from choosing a wrong alias frequency.

The absence of our signal $f_3$ in their data may be due to a similar
reason. Two out of their three yearly data sets spanned two weeks, and one
consisted two single-week runs spread by 25 days. The beating phenomenon
of $f_3$ with either $f_1$ or $f_2$ is therefore never properly resolved.
Due to this particular unfortunate time distribution $f_3$ can be
suppressed below detectability, as we have checked with numerical
simulations. It is therefore well possible that our $f_3$ was present in
their data but could not be detected. We note that the two combination
frequencies detected in our data may also be present in the measurements
by Heynderickx \& Haug (1994), but if so, their amplitude was lower.

Finally, we would like to point out that the lowest panel of Fig.\,6 also
shows a residual mound of power, even more pronounced than in the residual
periodogram of our data. The combined, scaled and average UBV residuals 
of the data by Heynderickx \& Haug (1994) have an rms scatter of 9.8 mmag,
also much higher than their measurement accuracy. As these authors used a
comparison star different from ours, we strengthen the conclusion that
there are unresolved frequencies in the light curves of IL Vel.

\subsection{Variable comparison stars}

\subsubsection{HD 90434}

This was the original local comparison star for V433 Car. When we examined
amplitude spectra of the differential data reduced during the course of
the observing campaign, we noted some additional low-frequency variability
besides the pulsations of V433 Car. The differential light curves of HD
90434 relative to the comparison star for IL Vel, allowed us to identify
HD 90434 unambiguously as a new variable. Figure 7 contains the spectral
window and amplitude spectra of these data.

\begin{figure}
\includegraphics[width=99mm,viewport=5 00 305 299]{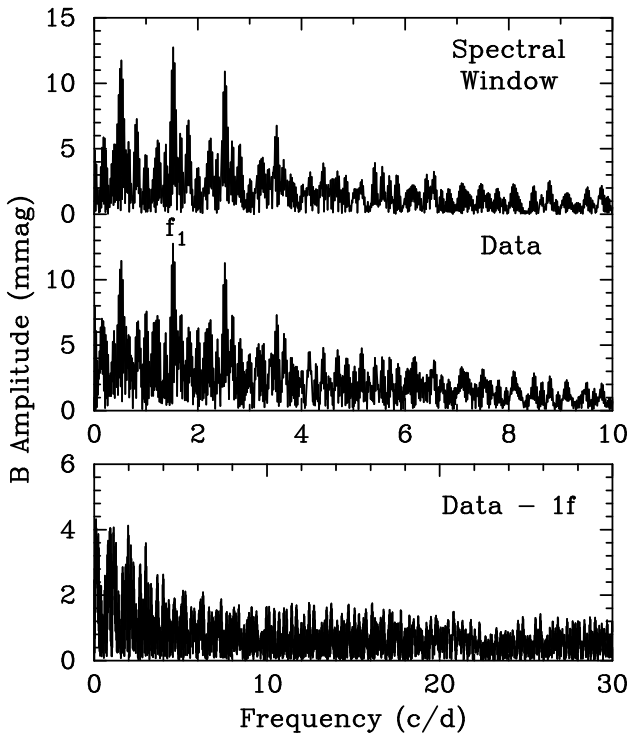}
\caption[]{Upper two panels: spectral window and amplitude spectrum of our
B filter data of HD 90434 relative to HD 79670. Lower panel: prewhitened
amplitude spectrum of the combined residual light curves out to the 
Nyquist frequency.}
\end{figure}

We detect one variation frequency for this star. Prewhitening it from the
data, combining the residual light curves as done for the $\beta$ Cephei
stars, and computing the amplitude spectrum of this data set results in
some evidence for additional variability (lowest panel of Fig.\,7), but
produces no more detections. The increase in noise level at low
frequencies may however also (partly) be due to residual effects of
extinction and sky transparency as the two stars are separated by about
10\degr in the sky. The frequency solution we derived for HD 90434 is
given in Table 5. We note that the amplitude of the variability increases
toward the blue, as confirmed by an examination for colour variations. No
significant phase shifts between the light curves in the different filters
were found.

\begin{table}
\caption[]{Results of the frequency analysis for the two comparison stars.
Frequencies are mean values from results of all four filters weighted by
the S/N ratio of the signal. Errors on the amplitudes were calculated with 
the formulae of Montgomery \& O'Donoghue (1999).The S/N ratio quoted is 
the average of the values in the individual filters.}
\begin{flushleft}
\begin{tabular}{ccc}
\hline
Star & HD 90434 & HD 89768 \\
\hline
Frequency (c/d) & 1.5229 $\pm$ 0.0012 & 0.6875 $\pm$ 0.0004\\
U Amplitude (mmag) & 15.9 $\pm$  1.2 & 13.6 $\pm$ 1.2 \\
B Amplitude (mmag) & 11.5 $\pm$  0.8 & 14.8 $\pm$ 0.6 \\
V Amplitude (mmag) & 9.2 $\pm$  0.7 & 12.7 $\pm$ 0.5 \\
v Amplitude (mmag) & 11.4 $\pm$  1.0 & 15.5 $\pm$ 0.9 \\
Mean S/N & 5.9 & 9.0 \\
\hline
\end{tabular}
\end{flushleft}
\end{table}

\subsubsection{HD 89768}

After discovery of its variability, HD 90434 was replaced as a comparison
star for V433 Car by HD 89768 whose spectral type of A1 IV/V places it
outside any known pulsational instability region, but we were no luckier
with this star. In fact, we only noticed that this object was variable as
well when it was too late to replace it with yet another star.

We again constructed differential light curves with respect to HD 79670
and show its Fourier analysis in the same fashion as the one in the
previous section in Fig.\,8. Again, a single frequency suffices to explain
the observed light variations, and some residual low-frequency variability
seems present that we attribute to atmospheric effects. Table 5 contains
our results for HD 89768 as well.

From the amplitudes in the different filters listed in Table 5 it appears
that this star also shows some colour variability. We have therefore
analysed the colour light curves as well, but we could not confirm this
suspicion. It seems that the different amplitudes in the four filters are
generated by noise effects and our errors are underestimates for
variations at these low frequencies.

\begin{figure}
\includegraphics[width=99mm,viewport=5 00 305 299]{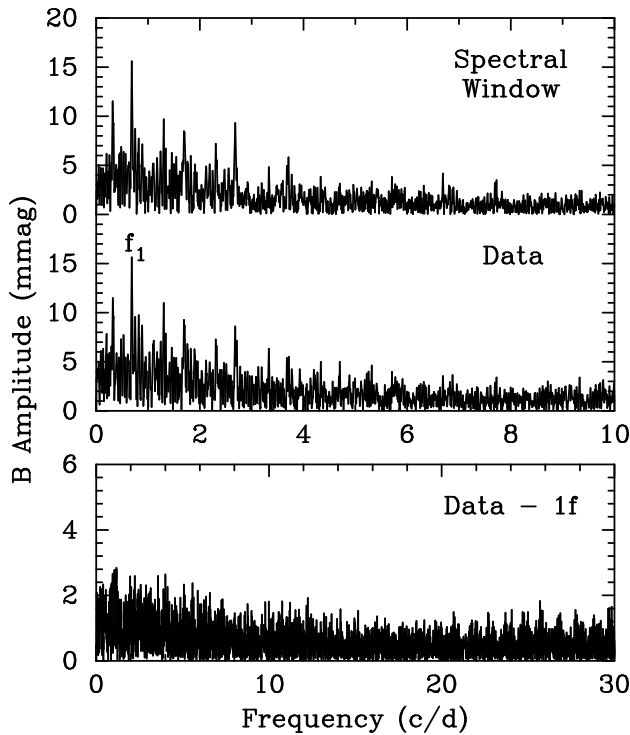}
\caption[]{Upper two panels: spectral window and amplitude spectrum of our
B filter data of HD 89768 relative to HD 79670. Lower panel: prewhitened
amplitude spectrum of the combined residual light curves out to the
Nyquist frequency.}
\end{figure}

\section{Basic stellar parameters}

\subsection{The $\beta$ Cephei stars}

The astrophysical analysis of our results is facilitated by firstly
restricting the physical parameter space occupied by these stars.
Consequently, we have estimated their masses and temperatures with
calibrations of multicolour photometry, as meaningful trigonometric
parallaxes are not available. Fortunately, all three stars have already
been measured in the Str{\o}mgren, Geneva and Johnson systems. We have
retrieved the corresponding standard values from the Lausanne-Geneva data
base ({\tt http://obswww.unige.ch/gcpd/gcpd.html}).

We have then applied different calibrations for effective temperature,
surface gravity and absolute magnitude to the three stars. It seems
important to use more than one calibration as their results often deviate.
We can then hope to get a good average result with a realistic
determination of its uncertainty. Absolute visual magnitudes were
calculated from the Str{\o}mgren $\beta$ index using Crawford's (1978)
results that also allow a determination of the interstellar reddening. We
find $A_v = 0.90$ for IL Vel, $A_v = 0.38$ for V433 Car and $A_v = 1.24$
for KZ Mus.

We used the formulae by Napiwotzki, Sch\"onberner \& Wenske (1993) to
calculate stellar temperatures from the Str{\o}mgren $[u-b]$ and $(b-y)_0$
indices as well as from Johnson $(B-V)_0$. The tables by Flower (1996)
allow another temperature estimate from $(B-V)_0$ and also provide
bolometric corrections. The latter were also determined from Drilling \&
Landolt's (2000) tables. Estimates of the surface gravities of the three
stars were obtained with the calibration by Smalley \& Dworetsky (1995)
from the Str{\o}mgren $\beta$ index, including a correction proposed by
Dziembowski \& Jerzykiewicz (1999). Finally, we applied the work by
K\"unzli et al. (1997) for the Geneva system that results in effective
temperatures and surface gravities. Average values for the resulting
effective temperatures and luminosities as well as error estimates are
given in Table 6.

\begin{table}
\caption[]{The adopted effective temperatures and luminosities for our
target stars from standard photometry.}
\begin{flushleft}
\begin{tabular}{lccc}
\hline
Parameter & IL Vel & V433 Car & KZ Mus \\
\hline
$T_{\rm eff}$ [kK] & 23.6 $\pm$ 0.6 & 26.6 $\pm$ 0.7 & 26.0 $\pm$ 0.7\\
log $L/L_{\sun}$ & 4.19 $\pm$ 0.22 & 4.20 $\pm$ 0.20 & 4.22 $\pm$ 0.20 \\
\hline
\end{tabular}
\end{flushleft}
\end{table}

\subsection{The variable comparison stars}

The HIPPARCOS parallax of HD 90434 (ESA 1997) has a relative error that is
too large to make it useful. The spectral classification and the
Str{\o}mgren colours of the star imply it is a late B star. Johnson and
Geneva colours of this object are also available at the Lausanne-Geneva
data base. Consequently, we applied the same calibrations that we have
already used for the $\beta$ Cephei stars to determine its effective
temperature and luminosity. We derive $T_{\rm eff}=12.2\pm0.4$\,kK and log
$L/L_{\sun} = 1.86\pm 0.10$.

No standard photometry is available for HD 89768. However, from the mean
magnitude differences of V433 Car from both of its comparison stars, we
can estimate $(B-V)=0.09$ and $(U-B)=0.04$ for HD 89768. With the
HIPPARCOS parallax of $2.09\pm0.69$\,mas and the galactic reddening law by
Chen et al. (1998), we can then estimate $(B-V)_0=0.01\pm0.03$ and
$M_v=-0.7\pm0.9$. From the calibration by Napiwotzki et al.\,(1993) we
then obtain $T_{\rm eff}=9250\pm400$\,K, and the luminosity results in log
$L/L_{\sun} = 2.3\pm 0.4$.

\subsubsection{The nature of HD 90434 and HD 89768}

The effective temperature and luminosity of HD 90434 as derived above put
it into the instability region of the slowly pulsating B (SPB) stars
(Dziembowski, Moskalik \& Pamyatnykh 1993). The colour variability we
found combined with the variation being in phase in the different filters
supports this hypothesis and argues against an interpretation in terms of
rotational modulation of a chemically peculiar star; the Str{\o}mgren
$m_1$ index for HD 90434 does not suggest it is an Ap star either.

On the other hand, HD 89768 is not located in any known pulsational
instability region, and we could not detect colour variability in its
light curves. We suspect that the variability may be binary-induced, but
spectroscopic data would be required to confirm our hypothesis.

\section{Mode identification}

The ranges in effective temperature and luminosity of the $\beta$ Cephei
stars listed in Table 6 are an essential ingredient for mode
identification. We have followed the method proposed by Balona \& Evers
(1999, see their paper for a detailed description). It uses theoretically
calculated nonadiabatic parameters to determine the amplitude ratios
between different wavebands that are the discriminator between the
different modes of pulsation.

We computed stellar evolutionary models by means of the Warsaw-New Jersey
evolution and pulsation code (described, for instance, by Pamyatnykh et
al. 1998) for solar chemical composition. The models spanned a mass range
between $8-16 M_{\sun}$ in steps of $1 M_{\sun}$. Guided by the projected
rotational velocities of the stars listed in Table 1, we chose a
rotational velocity of 100 km/s on the ZAMS for models representing IL Vel
and KZ Mus, but 260 km/s for V433 Car. By comparing the model evolutionary
tracks with the stellar parameters in Table 6, we inferred that IL Vel is
a star with a mass between $10.5-13 M_{\sun}$ in the second half of its
main-sequence evolution, whereas KZ Mus and V433 Car are less evolved but
somewhat more massive ($11.5 M_{\sun} < M_{\ast} < 14 M_{\sun}$).

We then proceeded by calculating theoretical UvBV amplitudes for all
pulsationally unstable modes along model sequences that spanned the
physical parameter space of our targets, whereby we extended the error
bars in Table 6 by up to a factor of 2 to be conservative. We did not
consider the phase shifts between the different photometric bands because
their measured relative error is much higher than that for amplitude
ratios. We have not found phase shifts significant at the 3$\sigma$ level
and they have therefore no discriminative power.

The resulting theoretical amplitudes in the four passbands were normalised
to that in the U filter, and the average amplitude ratio and its rms error
were calculated. In this way we can also get an idea of the uncertainties
involved in the calculation of theoretical results that need to be
considered when being matched with the observations. 

The comparison between the theoretical and observed colour amplitude
ratios of IL Vel is shown in Fig.\,9. We also include the UBV colour
amplitude ratios from the data by Heynderickx \& Haug (1984) with the
correct frequencies $f_1$ and $f_2$.

\begin{figure}
\includegraphics[width=99mm,viewport=5 00 305 415]{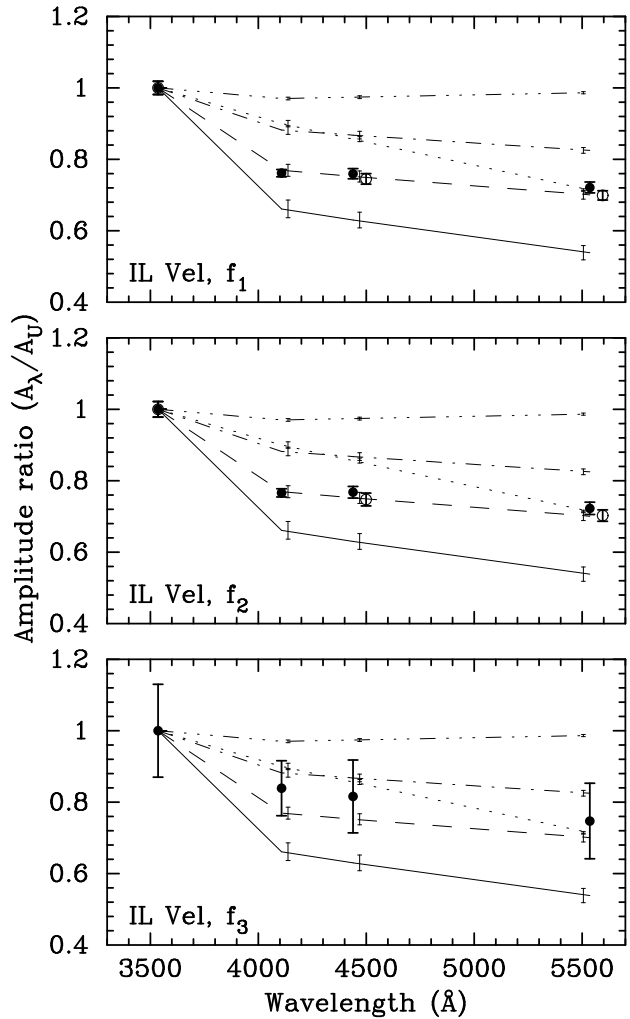}
\caption[]{Observed and theoretical UvBV amplitude ratios (lines) for
$0\leq\ell\leq4$ for IL Vel. Amplitudes are normalised at U. The filled
circles with error bars are from our measurements, the open circles
(somewhat shifted in wavelength for reasons of display) with error bars
are for the data by Heynderickx \& Haug (1984). The full lines are for
radial modes, the dashed lines for dipole modes, the dashed-dotted lines
for quadrupole modes, the dotted lines for modes of $\ell=3$ and the
dashed-dot-dot-dotted lines are for $\ell=4$. The small error bars, also
slightly shifted in wavelength, denote the uncertainty in the theoretical
amplitude ratios. The upper panel is for mode $f_1$, the middle one for
$f_2$, and the lower one for $f_3$.}
\end{figure}

The identification for the two strongest modes of IL Vel is unambiguous:
both have a spherical degree $\ell=1$. The third pulsation mode we found
that has a frequency intermediate between the other two is also nonradial
and could be $\ell=1, 2$ or 3. Our mode identification is therefore
consistent with that by Heynderickx et al. (1994), although they have used
an incorrect alias and spurious frequencies.

Observed and theoretical amplitude ratios for the four modes of V433 Car
are plotted in Fig.\,10. The identifications we derive from that figure
are not as clear as the ones for IL Vel. Mode $f_1$ is most likely
$\ell=2$, and the amplitude ratios for $f_2$ agree best with $\ell=1$.
All we can say about mode $f_3$ is that it is
nonradial ($\ell=2$ or 4?), and mode $f_4$ may be $\ell=0$, 1, 2 or 3.

Heynderickx et al. (1994) arrived at similar conclusions concerning the
identification of the modes of V433 Car, and they also reported
inconsistencies of their identifications in different photometric
systems. We must therefore consider the attempts to derive mode
identifications for V433 Car as mostly inconclusive.

\begin{figure}
\includegraphics[width=99mm,viewport=5 00 305 545]{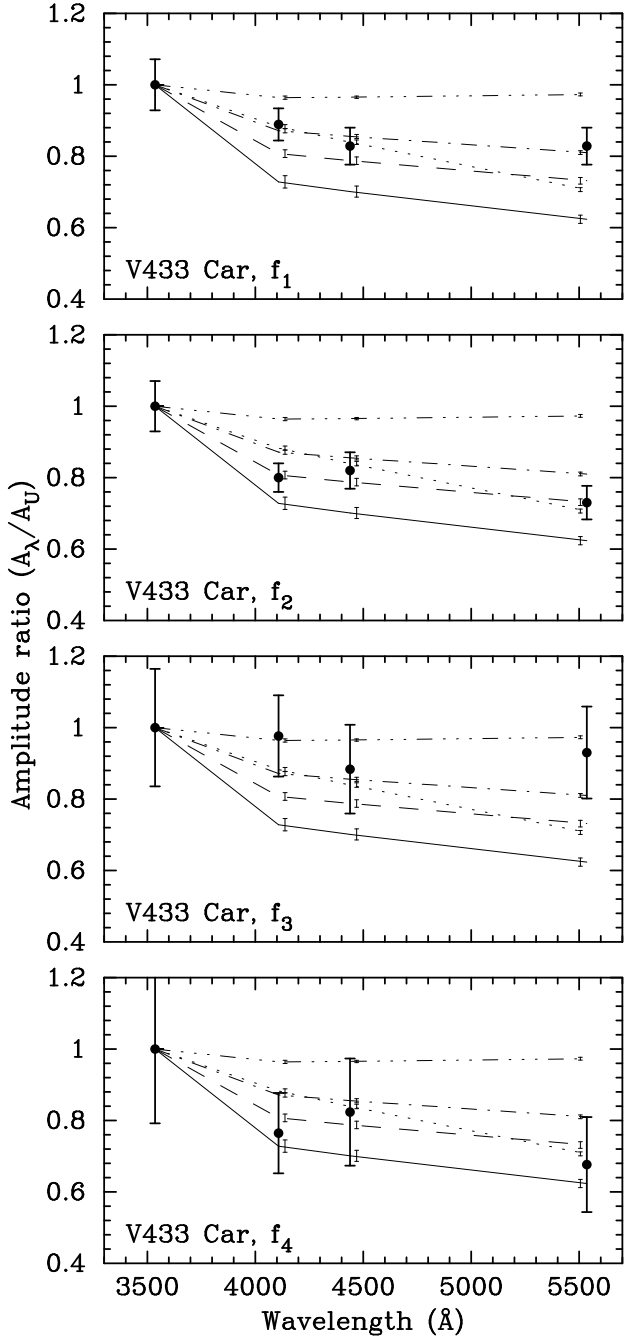} 
\caption[]{Observed (filled circles with error bars) and theoretical UvBV
amplitude ratios (lines) for $0\leq\ell\leq4$ for V433 Car. Amplitudes are
normalised at U. The full lines are for radial modes, the dashed lines for
dipole modes, the dashed-dotted lines for quadrupole modes, the dotted
lines for modes of $\ell=3$ and the dashed-dot-dot-dotted lines are for
$\ell=4$. The small error bars somewhat shifted in wavelength denote the
uncertainty in the theoretical amplitude ratios. The upper panel is for
mode $f_1$, the second for $f_2$, the third for $f_3$, and the lower one
for $f_4$.}
\end{figure}

Turning now to KZ Mus, we show the comparison of its observed amplitude
ratios with their theoretical predictions in Fig.\,11. The identifications
for the highest-amplitude modes of this star are very clear. The strongest
mode of KZ Mus is quadrupole ($\ell=2$), whereas mode $f_2$ is radial.
This is in excellent agreement with the identification by Aerts (2000).

As we have a much larger data set available than Aerts (2000) did, we can
also provide an identification for the third mode of KZ Mus, which is a
dipole ($\ell=1$). Only the identification for the weak fourth mode of this
star does not point to an unambiguous spherical degree, but we can safely
that it is nonradial with $\ell \leq 3$; a radial identification is ruled 
out because this mode's frequency is too close to that of the radial mode $f_2$.

\begin{figure}
\includegraphics[width=99mm,viewport=5 00 305 545]{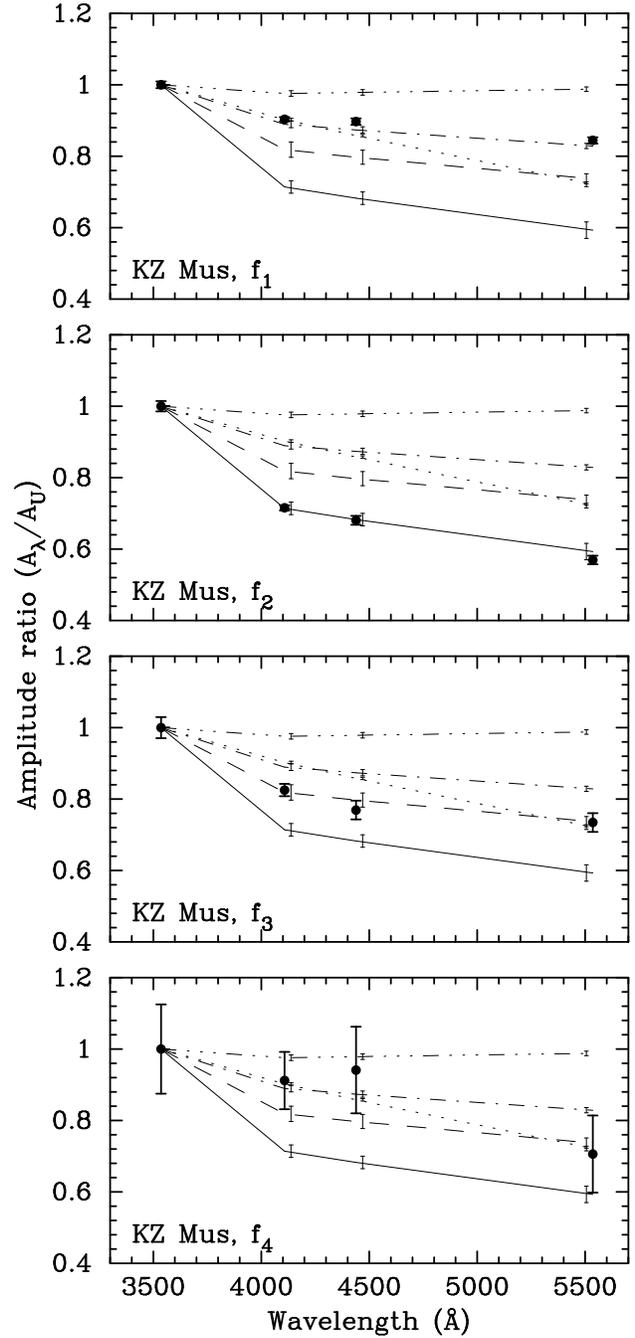} 
\caption[]{Observed (filled circles with error bars) and theoretical UvBV
amplitude ratios (lines) for $0\leq\ell\leq4$ for KZ Mus. Amplitudes are
normalised at U. The full lines are for radial modes, the dashed lines for
dipole modes, the dashed-dotted lines for quadrupole modes, the dotted
lines for modes of $\ell=3$ and the dashed-dot-dot-dotted lines are for
$\ell=4$. The small error bars somewhat shifted in wavelength denote the
uncertainty in the theoretical amplitude ratios. The upper panel is for
mode $f_1$, the second for $f_2$, the third for $f_3$, and the lower one
for $f_4$.}
\end{figure}

Finally, we also look for a possible mode identification for our SPB
candidate HD 90434. For this star, we adopted 3$M_{\sun}$ models near the
ZAMS guided by the results from Sect. 4.2 to determine the theoretical
UvBV amplitude ratios. We then proceeded in the same manner as for the
$\beta$ Cephei stars, and present the comparison between theoretical and
observed amplitude ratios in Fig.\,12.

\begin{figure}
\includegraphics[width=99mm,viewport=3 00 300 155]{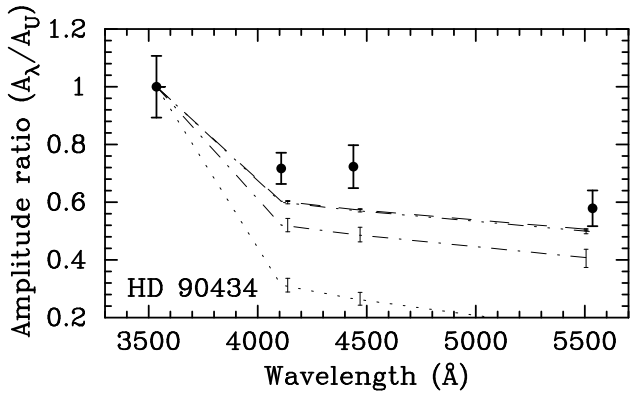} 
\caption[]{Observed (filled circles with error bars) and theoretical UvBV 
amplitude ratios (lines) for $1\leq\ell\leq4$ for HD 90434. Amplitudes are
normalised at U. The dashed lines are for dipole modes, the dashed-dotted
lines for quadrupole modes, the dotted lines for modes of $\ell=3$ and the
dashed-dot-dot-dotted lines are for $\ell=4$. The small error bars
somewhat shifted in wavelength denote the uncertainty in the
theoretical amplitude ratios.}
\end{figure}

Modes of spherical degrees $\ell=1$ and 4 give the best agreement, but
$\ell=2$ also seems a viable identification. The long period of the star
and the normalisation of the amplitudes in the U band (where the absolute
error on the amplitude is largest) are suspected to result in larger
systematic contributions to the observational errors than the formal
values we adopted would include. Taking the effects of geometrical
cancellation (Dziembowski 1977) into account, the most likely reason for
the variability of HD 90434 is pulsation in a dipole mode.

\section{Asteroseismology}

\subsection{IL Vel}

We will now take advantage of the information gathered in the previous
sections to understand the pulsational behaviour of our three target stars
by means of comparison with model calculations. We have again used the
Warsaw-New Jersey stellar evolution and pulsation code for this part of
the analysis. We note that we do not attempt an exhaustive seismic
analysis of the stars because too few identified pulsation modes are
available for such purposes. Instead, we want to determine what we can
learn about the stars by means of standard models and procedures and we 
want to assess the potential for asteroseismology of $\beta$ Cephei stars.

One tool that has been established for constraining the range of physical
parameters of the seismological model of a star is stability analysis (see
Pamyatnykh 2003 for a review). As a model of a pulsating star evolves, its
radius changes, and the range of pulsationally unstable radial overtones
of the eigenfrequencies varies as well. In conclusion, the frequency range
of the unstable eigenmodes depends on the location of the model in the
instability strip. This frequency range can then be matched with the one
actually observed in the star.

Applying this method to IL Vel, we can take advantage of our mode
identification. As all the modes for which we have a certain
identification are $\ell=1$, we can restrict the comparison to this single
spherical degree only. Figure 13 shows evolutionary tracks for models
between 8 and 16\,$M_{\sun}$ (we assumed solar chemical composition and a
rotational velocity of 100 km/s on the ZAMS) in the theoretical HR diagram
together with the parameters of the star from Table 6; models with a
matching range of unstable frequencies are indicated.

\begin{figure}
\includegraphics[width=99mm,viewport=5 00 305 255]{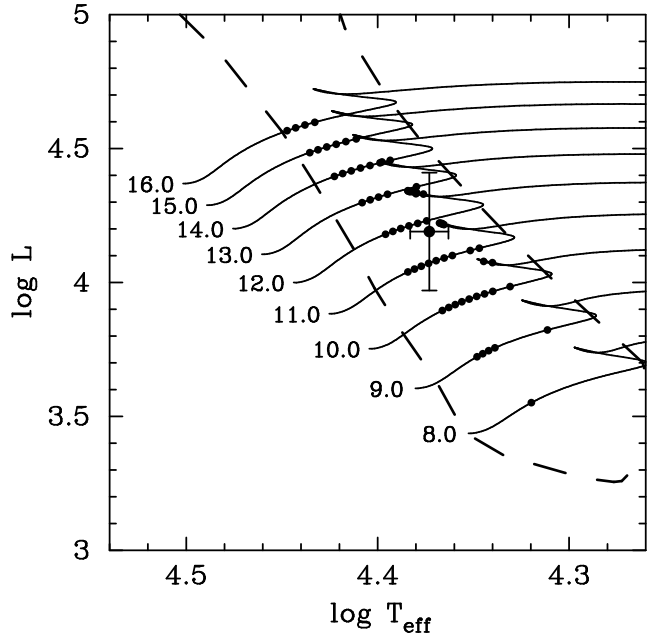}
\caption[]{The position of IL Vel in the theoretical HR diagram. Some
stellar model evolutionary tracks labeled with their masses (full lines)
and the theoretical borders of the $\beta$ Cephei star instability strip
(Pamyatnykh 1999, dashed lines) are included for comparison. The sections
of the evolutionary tracks that are marked with filled circles represent
models in which the observed frequency range of the star is unstable.}
\end{figure}  

\begin{figure}
\includegraphics[width=96mm,viewport=02 00 292 506]{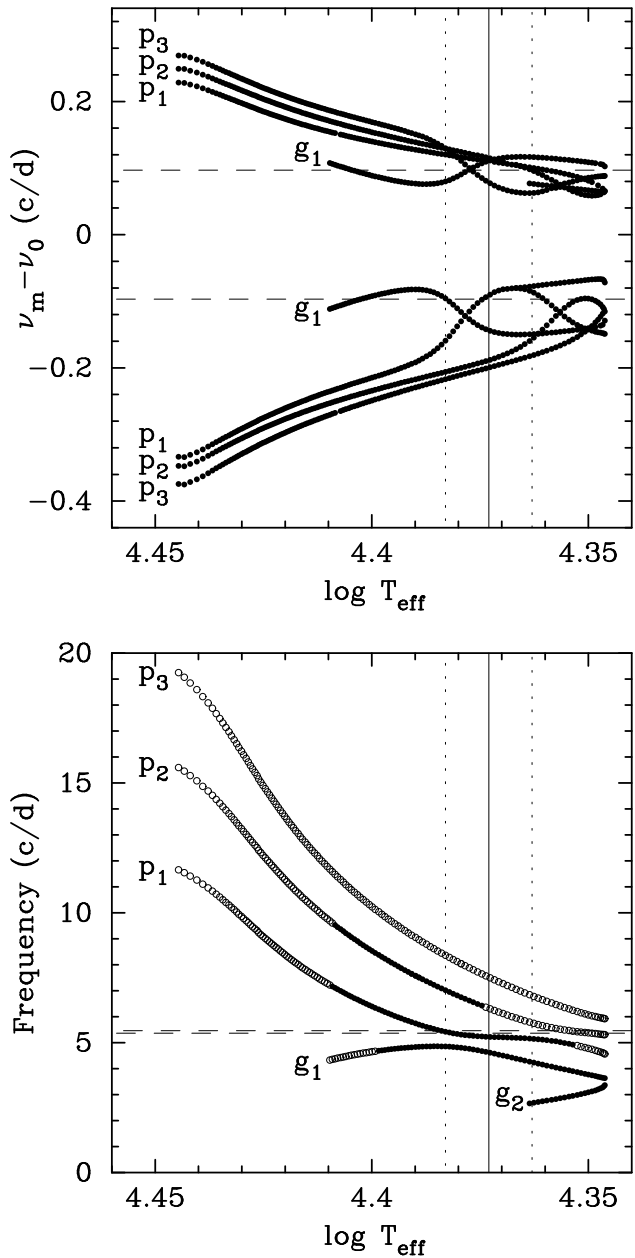}

\caption[]{Upper panel: the size of rotational splitting for $\ell=1$
modes of a 12 $M_{\sun}$ model with $v_{\rm rot,ZAMS}=70$\,km/s along its
evolution along the main sequence. The first three p-modes and the first
g-mode are indicated. The horizontal dashed lines denote the frequency
difference between the two strongest modes of IL Vel, and the vertical
solid line with parallel dotted lines is the effective temperature of the
star and its error estimate from Table 6. Lower panel: stability and
frequency evolution of $\ell=1,m=0$ mode frequencies in the same model.
Modes that are pulsationally unstable are denoted with filled circles, the
open circles are for stable modes. The two dashed lines represent the
observed $\ell=1$ frequencies, and the vertical lines again labels the
effective temperature of the star and its error estimate. Note the avoided 
crossing of the $p_1$ and the $g_1$ mode starting at about log $T_{\rm 
eff}=4.39$.}
\end{figure}  

Inspection of Fig.\,13 shows very good agreement between the star's
position in the HRD and the range of models that have the same frequency
domain unstable. We note that the sequences of matching models sometimes
have gaps as they evolve along the main sequence. This is due to a change
in the type of modes that produce the match and their frequency evolution
which is affected by avoided crossings. We caution that the strength of
pulsational driving in $\beta$ Cephei stars is strongly dependent on metal
abundance (Moskalik \& Dziembowski 1992, Pamyatnykh 1999). Therefore the
results in Fig.\,13 are only an indication of the range of possible models
for IL Vel.

The two $\ell=1$ modes of IL Vel have similar frequencies; they are spaced
by only 0.0986 c/d. The projected rotational velocity of the star ($v \sin
i = 65$\,km/s, cf. Table 1) implies a rotational period $P_{\rm rot} <
5.8$\,d for the star assuming a mass of 12 $M_{\sun}$ as indicated by
Fig.\,13. At first glance, this suggests that those two modes cannot be
caused by rotational splitting of a single mode. Three explanations for
this close splitting may be possible. First, the star's pulsations are
just undergoing an avoided crossing phenomenon (e.g. see Aizenman, Smeyers
\& Weigert 1976) resulting in a close proximity of a p- and a g-mode.
Second, the two $\ell=1$ modes are the central and prograde components of
a multiplet, and the second-order rotational splitting is large enough to
produce such a small frequency difference. Third, we may observe a gravity
mode; the first-order rotational splitting of such a mode is approximately
$\nu\approx (1-(\ell(\ell+1))^{-1})P_{\rm rot}^{-1}$ (Winget et al. 1991).

We examine these hypotheses in Fig.\,14, where we show the evolution of
the rotational $m$-mode splitting and the frequencies of the $\ell=1$
modes of a 12 $M_{\sun}$ model with $v_{\rm rot,ZAMS} = 70$\,km/s along
the main sequence. The first three pressure (p) and the first two gravity
(g) modes are shown. From the upper panel of this figure it becomes clear
that the observed frequency splitting of IL Vel can be easily reproduced
by the model in the corresponding temperature range. In fact, this would
be possible with models rotating as fast as 110 km/s. On the other hand,
the rotational frequency splitting of any of these modes never becomes
small enough that the observed frequency difference can be due to $\ell=1$
modes of the same type and radial overtone but with $m=1$ and $m=-1$. This
also implies that the third mode of IL Vel cannot be another member of
this multiplet.

To determine the modes that are potentially responsible for the pulsations
of IL Vel, we have traced the frequencies of the individual $\ell=1$ modes
of the same model as used before along the main sequence in the lower
panel of Fig.\,14. The mode originating as $p_1$ on the ZAMS has a
frequency very similar to the two observed for IL Vel over the whole
temperature range estimated for the star, and is unstable as well. This
mode has already undergone an avoided crossing with the $g_1$ mode and
should therefore also have gravity mode characteristics, i.e. it is a
mixed mode. In higher-mass models, dipole modes at this frequency become
stable against pulsations, and in models below 11\,$M_{\sun}$ the $g_1$
mode reproduces the observed frequency in the temperature range
appropriate for IL Vel.

Another observation from Fig.\,14 is that the $p_1$ and $g_1$ modes never
come closer together than about 0.5 c/d. It is therefore unlikely that the
two modes observed in IL Vel originate from different modes that just
happen to undergo an avoided crossing. If we consider the hypothesis that
the proximity of the two observed $\ell=1$ modes is caused by the effects
of second-order rotational splitting, we find that $v_{\rm rot} \simgt
160$ km/s, and hence $i < 25\degr$, which is also unlikely. In addition,
the high amplitude of the prograde mode and the absence of the retrograde 
component of the hypothesised triplet would also be difficult to explain.

We conclude that the two dipole pulsation modes of IL Vel are most likely
due to rotationally split components of the mode originating as $p_1$ on
the ZAMS. One of these modes is $m=0$, the other one $|m|=1$, and 65 km/s 
$< v_{\rm rot} <$ 110 km/s.

\subsection{V433 Car}

We have performed a similar analysis for V433 Car. Regrettably, we could
not obtain unambiguous $\ell$ identifications for any of its modes. On the
other hand, the excited frequency range of the star is larger, which may
however be related to it being a fast rotator. In any case, we show the
match of unstable model frequencies to that of the observed ones in
Fig.\,15. The model evolutionary tracks are again for solar chemical
composition, but for a rotational velocity of 260 km/s on the ZAMS, and we
considered all modes with $0 \leq \ell \leq 4$.

\begin{figure}
\includegraphics[width=99mm,viewport=5 00 305 255]{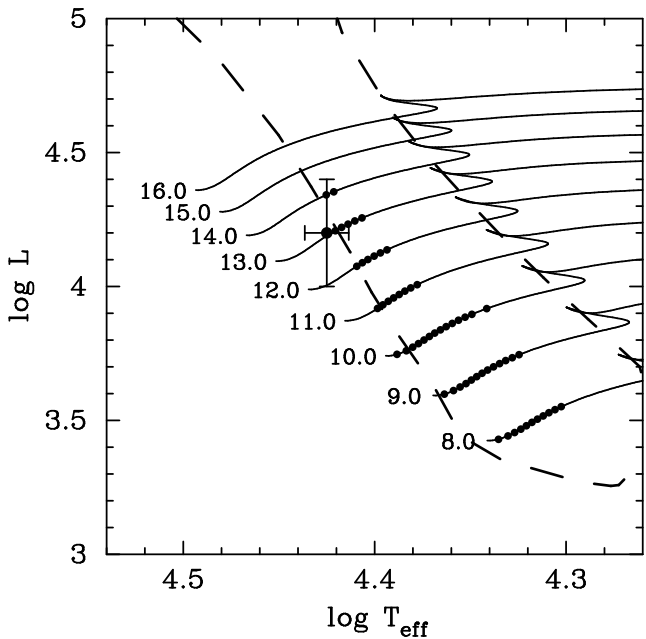}
\caption[]{The position of V433 Car in the theoretical HR diagram.
Some stellar model evolutionary tracks labeled with their masses (full
lines) and the theoretical borders of the $\beta$ Cephei star instability
strip (Pamyatnykh 1999, dashed lines) are included for comparison. The
sections of the evolutionary tracks that are marked with filled circles
represent models in which the observed frequency range of the star is
unstable. For some models, we obtain unstable pulsation modes somewhat
outside the blue edge of the theoretical instability strip, which is due
to $\ell=4$ modes; the borders indicated are only for $\ell \leq 2$.}
\end{figure}  

Again, the agreement between theory and observation is fairly good, even
when keeping in mind that our analysis does not take the frequency spread
due to fast rotation of the star. However, as the observational results
only allow us a rough qualitative comparison between observations and
theory for V433 Car, this is not regarded as being important. V433 Car
appears to be an object of about 13 $M_{\sun}$ that has just entered the
$\beta$ Cephei star instability strip.

\subsection{KZ Mus}

The asteroseismological prospects of KZ Mus appear much better. In
particular, we have unambiguously identified one of its modes as radial,
and two more modes also have unique $\ell$ assignments. In the same 
fashion as in the previous sections, we show model evolutionary tracks
(for $v_{\rm rot, ZAMS}=100$\,km/s) in the theoretical HR diagram, the
star's position in it, and models with unstable modes in the observed
frequency range in Fig.\,16. In addition, we have indicated the locations
of models that have a radial mode at 5.9506 c/d as observed for KZ Mus.

\begin{figure}
\includegraphics[width=99mm,viewport=5 00 305 255]{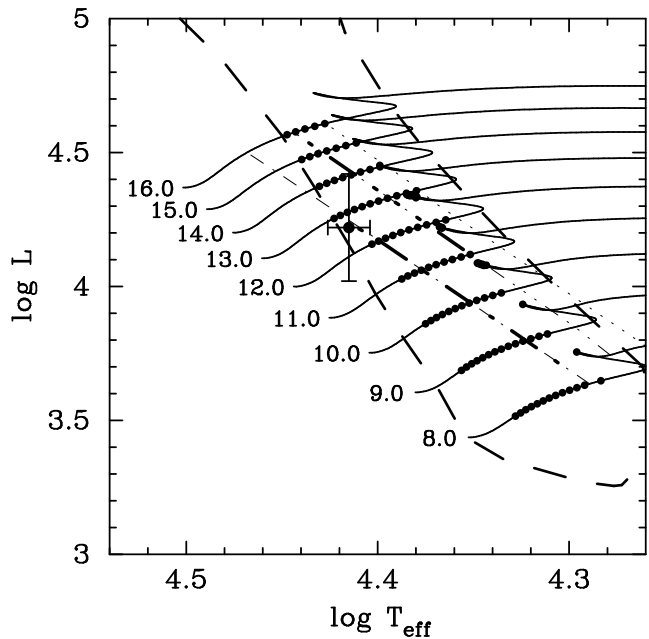}
\caption[]{The position of KZ Mus in the theoretical HR diagram. Some
stellar evolutionary tracks labeled with their masses (full lines) 
and the theoretical borders of the $\beta$ Cephei star instability strip 
(Pamyatnykh 1999, dashed lines) are included for comparison. The
sections of the evolutionary tracks that are marked with filled circles   
represent models in which the observed frequency range of the star is 
unstable. The lines going from the top left to the bottom right are
locations of models that have a radial mode at 5.9506 c/d. For models on
the dashed-dotted line this would be the fundamental radial mode, those
connected by the dotted line show the first overtone at this frequency,
and the ones connected by the dashed-dot-dot-dotted line have their second
radial overtone at 5.9506 c/d. The thick parts of these lines connect
the models which are pulsationally unstable.}
\end{figure}

As for the other two stars, we find no disagreement between models with
unstable modes in the observed frequency range and the observationally
determined position of the star in the HRD. Figure 16 also allows us to
reject the second overtone as the identification for the radial mode of KZ
Mus, as it is inconsistent with the star's evolutionary state, and it
never becomes pulsationally unstable.

However, we cannot say at this point whether the radial mode is the
fundamental or first overtone. Models with the fundamental radial mode at
5.9506 c/d are in better agreement with the stellar parameters from Table
6, but it may be stable in models with the mass inferred for the star. On
the other hand, for this mode to be the first overtone, only models at the
upper limits on mass and luminosity seem appropriate.

We attempted to find further constraints on the parameter space occupied
by KZ Mus by using the nonradial modes that we identified unambiguously;
any model representing the star should have modes of the same $\ell$ in
the correct frequency range.  All models with the radial fundamental or
first overtone mode at 5.9506 c/d that we investigated could be used to
reproduce the observed nonradial modes given the unknown angular
rotational velocity of the star. However, when imposing mode stability as
an additional criterion, it is found that model sequences of 12 and 13
$M_{\sun}$ give best agreement with the observations.

In this case, it is most likely that $f_2$ is the radial fundamental mode
of KZ Mus. Then the $\ell=1$ mode would be associated with the mode
originating as $p_1$ at the ZAMS and the observed $\ell=2$ most likely
corresponds to the mode originating as $g_1$.

\section{Summary, discussion and conclusions}

In an attempt to investigate the asteroseismological potential of the
$\beta$ Cephei stars, we have obtained 127 -- 150 h of differential
photoelectric photometry for each of the three pulsators IL Vel, V433 Car
and KZ Mus. In line with the recent results by Stankov et al. (2002) and
Cuypers et al. (2002), we have shown that low-amplitude pulsation modes
are present in $\beta$ Cephei stars, and that large observational efforts
to determine their mode spectra in detail are justified. We therefore
confirm the importance of $\beta$ Cephei stars as asteroseismological
targets for both ground and space based measurements.

By using the amplitude ratios of the different pulsation modes in the UvBV
filters of the Johnson and Str{\o}mgren systems, we performed mode
identification. We obtained unambiguous results for all pulsation modes
with photometric amplitudes larger than about 10 mmag in V, or with
relative errors on their amplitude smaller than 3\%. The modes we could
identify were the two dominant ones of IL Vel and the three strongest of
KZ Mus, but none of V433 Car.

This may not only be due to its photometric amplitudes that are small
compared to those of the other two $\beta$ Cephei stars. We have shown
that V433 Car is a fast rotator, and therefore mode coupling (see
Daszy{\'n}ska-Daszkiewicz et al. 2002 and references therein) may occur.
Mode coupling affects the predicted colour amplitude ratios and phase
shifts. Since we have not taken this into account in our analysis, it may
be at least partly responsible for the ambiguous mode identifications for
this star.

On the other hand, our results for the more slowly rotating stars appear
reliable, which represents a good motivation for studying a larger number
of multiperiodic $\beta$ Cephei stars in the future. Our limit of the
relative error of 3\% of the photometric amplitudes for successful mode
identification can then serve as a guideline for planning future
campaigns.

We have performed model calculations for our three targets. The outcome
was also quite encouraging. Although we only detected up to four modes per
star (thus our obervational constraints on the models are not very
strong), in no case disagreements between the observational results and
model predictions were found. The observationally determined positions of
the stars in the HR diagram could be reproduced by models that have the
same range in pulsation frequencies unstable as are excited in the real
stars.

In addition, we were able to place constraints on the types of mode
observed in IL Vel and KZ Mus. We have shown that the most likely
explanation for the two dipole modes of IL Vel are rotationally split
components (one necessarily being $m=0$) of the mode originating as $p_1$
on the ZAMS. The third mode of IL Vel cannot be another multiplet member,
and the equatorial rotational velocity of the star must be smaller than
110 km/s. If the radial mode observed in KZ Mus is the fundamental, then
the high-amplitude $\ell=2$ mode is most likely the one originating as
$g_1$, and the $\ell=1$ mode we identified is probably the one originating
as $p_1$.

We cannot push the asteroseismological analysis of the three stars further
at this point because too few modes have been detected and identified for
detailed modelling. However, we strongly suspect that more modes are
present in IL Vel that just need better time resolution to be detected. KZ
Mus may have more modes as well, and a unique mode identification for its
fourth mode may place a constraint on the star's rotation frequency.
Further work on both stars should also include time-resolved spectroscopic
observations to determine the azimuthal order $m$ of the high-amplitude
nonradial modes; although both objects are $V \approx 9.1$, the spectra we
obtained suggests this is possible. Determinations of the metal abundances
of the stars would also be a valuable ingredient for model calculations.
With all these observations in hand, precision asteroseismological studies
of IL Vel and KZ Mus appear possible.

\section*{ACKNOWLEDGEMENTS}

This work has been supported by the Austrian Fonds zur F\"orderung der
wissenschaftlichen Forschung under grant R12 to GH and by travel support
for FR and TT under grant P14546-PHY. We thank Wolfgang Zima for helping
with some of the observations. GH is indebted to Alosha Pamyatnykh for
helpful discussions, for providing theoretical instability domains for the
$\beta$ Cephei stars and to him and Anamarija Stankov for comments on a
draft version of this paper. This research has made use of the SIMBAD and
VizieR databases, operated at CDS, Strasbourg, France and the GCPD
database, operated at the Institute of Astronomy of the University of
Lausanne.

\end{document}